\input harvmac
 \def\quad{{\ \ }}
\def\ZZ{{\bf Z}}
\def\DD{{\bf D}}

\def\ra{\r\input{unoriented.tex}
ightarrow}
\def\frac#1#2{{{#1}\over {#2}}}

\def\ra{\rangle}


\def\sym{  \> {\vcenter  {\vbox
                  {\hrule height.6pt
                   \hbox {\vrule width.6pt  height5pt
                          \kern5pt
                          \vrule width.6pt  height5pt
                          \kern5pt
                          \vrule width.6pt height5pt}
                   \hrule height.6pt}
                             }
                  } \>
               }
\def\fund{  \> {\vcenter  {\vbox
                  {\hrule height.6pt
                   \hbox {\vrule width.6pt  height5pt
                          \kern5pt
                          \vrule width.6pt  height5pt }
                   \hrule height.6pt}
                             }
                       } \>
               }
\def\anti{ \>  {\vcenter  {\vbox
                  {\hrule height.6pt
                   \hbox {\vrule width.6pt  height5pt
                          \kern5pt
                          \vrule width.6pt  height5pt }
                   \hrule height.6pt
                   \hbox {\vrule width.6pt  height5pt
                          \kern5pt
                          \vrule width.6pt  height5pt }
                   \hrule height.6pt}
                             }
                  } \>
               }

 \lref\SeibergBX{
  N.~Seiberg,
  ``Observations on the moduli space of two dimensional string theory,''
  JHEP {\bf 0503}, 010 (2005)
  [arXiv:hep-th/0502156].
}

 \lref\GomisVI{
  J.~Gomis and A.~Kapustin,
  ``Two-dimensional unoriented strings and matrix models,''
  JHEP {\bf 0406}, 002 (2004)
  [arXiv:hep-th/0310195].
}
 
\Title{\vbox{\baselineskip12pt\hbox{}\hbox{}}}
{\vbox{\centerline{
Anomaly Cancellation in Noncritical String Theory}}}

\centerline{Jaume Gomis\foot{jgomis@perimeterinstitute.ca}}
\medskip\medskip

\bigskip\centerline{\it Perimeter Institute for Theoretical Physics}
\centerline{\it Waterloo, Ontario N2L 2Y5, Canada}
\vskip .3in

\centerline{Abstract}

We construct new two dimensional unoriented superstring theories in two dimensions with a  chiral closed string spectrum and show that anomalies cancel upon supplying  the appropriate chiral open string degrees of freedom imposed by tadpole cancellation.

\Date{8/2005}

\newsec{Introduction and Conclusion}

The ``miraculous"  cancellation of anomalies in ten dimensional superstring theory 
\lref\GreenSG{
  M.~B.~Green and J.~H.~Schwarz,
  ``Anomaly Cancellation In Supersymmetric D=10 Gauge Theory And Superstring
  Phys.\ Lett.\ B {\bf 149}, 117 (1984).
}
\GreenSG\  gave rise for the first time to a consistent framework that incorporates quantum gravity and chiral gauge couplings akin to the ones in the standard model. By now, a beautiful picture has emerged of how potential anomalies can cancel upon compactification from ten dimensions.

In this note we construct new two dimensional non-critical superstring theories with parity violating couplings and show that anomalies can be cancelled. These low dimensional theories are simple toy models of string theory which nevertheless capture -- in  a calculable setting -- important concepts and techniques of more realistic string vacua. These include
D-branes, RR fluxes, holography, dualities  and as we show in this note cancellation of gauge and gravitational anomalies.

We construct these theories by performing suitable orientifold projections of the known two dimensional noncritical superstring theories. It is shown that theories for which the orientifold plane carries RR charge have an anomalous closed string spectrum. Nevertheless, we show that imposing RR tadpole cancellation -- a basic worldsheet consistency condition
\lref\PolchinskiTU{
  J.~Polchinski and Y.~Cai,
  ``Consistency Of Open Superstring Theories,''
  Nucl.\ Phys.\ B {\bf 296}, 91 (1988).
}
\lref\CallanWZ{
  C.~G.~.~Callan, C.~Lovelace, C.~R.~Nappi and S.~A.~Yost,
  ``Loop Corrections To Superstring Equations Of Motion,''
  Nucl.\ Phys.\ B {\bf 308}, 221 (1988).
}
\PolchinskiTU\CallanWZ\ -- introduces the correct chiral (open) string degrees of freedom to cancel all anomalies.

Some of these models have an unusual particle spectrum, consisting only of space-time fermions,  which makes it challenging to find a matrix model formulation of these theories; there are also no known unstable D-branes that can be used to define the closed string vacuum via tachyon condensation
\lref\SenII{
  A.~Sen,
  ``Stable non-BPS bound states of BPS D-branes,''
  JHEP {\bf 9808}, 010 (1998)
  [arXiv:hep-th/9805019].
}
\SenII.
 Recently, in 
\lref\HoravaTT{
  P.~Horava and C.~A.~Keeler,
  ``Noncritical M-theory in 2+1 dimensions as a nonrelativistic Fermi liquid,''
  arXiv:hep-th/0508024.
}
\HoravaTT, 
an M-theory formulation of noncritical string theory has been proposed. An important consistency check of that proposal is to identify the string models found in this note in the three dimensional M-theory construction.

The plan of the rest of the note is as follows. In section $2$ we summarize the known noncritical superstring theories in two dimensions, study their  symmetries  and classify the possible orientifold projections. In section $3$ the constraints imposed by anomaly cancellation and tadpole cancellation are presented. Section $4$ shows that the anomalies due to the chiral closed string spectrum of the orientifold models are cancelled by the open string degrees of freedom induced by tadpole cancellation.

\newsec{Two Dimensional Noncritical Superstring Theories}

In this section we summarize the known   superstring theories\foot{That is theories with ${\cal N}=(1,1)$ supersymmetry on the worldsheet. There are also heterotic theories 
with ${\cal N}=(1,0)$  supersymmetry 
\lref\McGuiganQP{
  M.~D.~McGuigan, C.~R.~Nappi and S.~A.~Yost,
  ``Charged black holes in two-dimensional string theory,''
  Nucl.\ Phys.\ B {\bf 375}, 421 (1992)
  [arXiv:hep-th/9111038].
}
\lref\DavisQE{
  J.~L.~Davis, F.~Larsen and N.~Seiberg,
  ``Heterotic strings in two dimensions and new stringy phase transitions,''
  arXiv:hep-th/0505081.
}
\McGuiganQP\DavisQE.} in two non-compact dimensions, their massless spectrum and 
  corresponding symmetries\foot{We assume throughout that we are in the pure linear dilaton vaccuum, i.e. we do not turn on any of the possible marginal deformations, which may break some of the symmetries.}.
We then classify the admissible  orientifold projections, which give rise to new unoriented string theories in two dimensions. The cancellation of anomalies and tadpoles  in these theories will be studied in the rest of the note.

Just like in ten dimensions, there are two consistent  ``Type II" 
\lref\SeibergBX{
  N.~Seiberg,
  ``Observations on the moduli space of two dimensional string theory,''
  JHEP {\bf 0503}, 010 (2005)
  [arXiv:hep-th/0502156].
}
\lref\GukovYP{
  S.~Gukov, T.~Takayanagi and N.~Toumbas,
  ``Flux backgrounds in 2D string theory,''
  JHEP {\bf 0403}, 017 (2004)
  [arXiv:hep-th/0312208].
}
\lref\TakayanagiGE{
  T.~Takayanagi,
  ``Comments on 2D type IIA string and matrix model,''
  JHEP {\bf 0411}, 030 (2004)
  [arXiv:hep-th/0408086].
}
\SeibergBX\GukovYP\TakayanagiGE\ GSO projections\foot{One can get isomorphic copies of these theories by the action of space-time parity, which reverses the sign of the GSO projection in all $R$ sectors.}:

\eqn\spectII{\eqalign{
\matrix{&\underline{\hbox{Type IIB}} &\underline{\hbox{Type IIA}}\cr
&  &\cr
&\hskip-20pt(NS+,NS+)&(NS+,NS+)\cr
&(NS-,R-)\rightarrow \Psi_-&\hskip+20pt(NS-,R+)\rightarrow \Psi_+\cr
&(R-,NS-)\rightarrow \tilde{\Psi}_-&\hskip+20pt(R-,NS-)\rightarrow \Psi_-\cr
&\hskip-8pt(R+,R+)\rightarrow C_+&\hskip-20pt (R+,R-).}}}
\medskip
\medskip
The level matching condition forces the GSO projection in the $(NS,R)$ and $(R,NS)$ sectors to be {\it different} than in ten dimensions. 

The spectrum of particles of the Type IIB string  is a left moving boson $C_+$ and two right moving Majorana-Weyl fermions $\Psi_-$ and $\tilde{\Psi}_-$, while the Type IIA string has a left  and right moving Majorana-Weyl fermion denoted by $\Psi_+$ and $\Psi_-$ respectively.

The discrete symmetry group of two dimensional   Type IIB string theory is  the dihedral group of four elements $\DD_4$, generated\foot{The relations in the group are   $a^4=b^2=1$ and $(ab)^2=1$.} by $a=\Omega(-1)^{F}$ and $b=\Omega$, where  $\Omega$ stands for the worldsheet parity operator and $(-1)^{F({\bar F})}$ denotes the left(right)-moving spacetime fermion number operator.  The complete list of elements of the symmetry group  is $G=\{1,\Omega, (-1)^F, (-1)^{\bar F}, (-1)^{F+{\bar F}},\Omega (-1)^F, \Omega (-1)^{\bar F},\Omega (-1)^{F+{\bar F}}\}$. The Type IIA discrete symmetry group is $\ZZ_2\times \ZZ_2$, generated by $(-1)^F$ and $(-1)^{\bar{F}}$. As in ten dimensions, the theory is not invariant\foot{If we relax the condition of translational invariance along the $x$ coordinate in which the dilaton does not vary, the symmetry group enhances to the dihedral group, which is now generated by $a=\Omega \hbox{R}(-1)^{F}$ and $b=\Omega \hbox{R}$, where $\hbox{R}$ is a space-time parity transformation acting by $x\rightarrow - x$.} under the action of $\Omega$.

There are two consistent Type 0   
\lref\TakayanagiSM{
  T.~Takayanagi and N.~Toumbas,
  ``A matrix model dual of type 0B string theory in two dimensions,''
  JHEP {\bf 0307}, 064 (2003)
  [arXiv:hep-th/0307083].
}
\lref\DouglasUP{
  M.~R.~Douglas, I.~R.~Klebanov, D.~Kutasov, J.~Maldacena, E.~Martinec and N.~Seiberg,
  ``A new hat for the c = 1 matrix model,''
  arXiv:hep-th/0307195.
}
\TakayanagiSM\DouglasUP\ GSO projections:

\eqn\specto{\eqalign{
\matrix{&\underline{\hbox{Type 0B}} &\underline{\hbox{Type 0A}}\cr
&  &\cr
&\hskip-20pt(NS+,NS+)&(NS+,NS+)\cr
&(NS-,NS-)\rightarrow T&\hskip+20pt(NS-,NS-)\rightarrow T\cr
&(R-,R-)\rightarrow C_-&\hskip-20pt(R-,R+)\cr
&\hskip-8pt(R+,R+)\rightarrow C_+&\hskip-20pt (R+,R-).}}}

Therefore, the spectrum of particles of the Type 0B string  is  two scalars $T$  and $C=C_++C_-$, while the Type 0A string has a single scalar field $T$.

The symmetry group of Type 0B string theory  is $\ZZ_2\times\ZZ_2\times \ZZ_2$, generated by $\Omega, (-1)^F$ and $(-1)^f$, where $(-1)^{f({\bar f})}$ denotes the left(right)-moving worldsheet fermion number operator. Type 0A string theory has $\DD_4$ as its\foot{If we allow for broken  translational invariance along $x$, the Type 0A string is  invariant under the action of $\hbox{R}$.}
group of symmetries, and is generated by $a=\Omega(-1)^f$ and $b=\Omega$.

\noindent

Now that we have found the symmetries of the various theories, we are ready to construct new unoriented superstring theories. The unoriented theories we construct are obtained by modding out a  string theory by the orientifold group $G=\{1,\Omega g\}$, where $\Omega g$ is a symmetry generated by the combined action of the worldsheet parity operator $\Omega$ and a discrete symmetry generator $g$. In order for the orientifold to be well defined we must find group elements $g$ such that $(\Omega g)^2=1$ when acting on the string  spectrum\foot{If $(\Omega g)^2\neq 1$, we can still construct a consistent orientifold group, but it is of the type $\tilde{G}=H_1\cup \Omega H_2$. This can be interpreted as an orientifold of an orbifold by $H_1$, but modding by $H_1$ yields another theory in the list, whose orientifolds we already considered. For example, in Type IIB, $(\Omega (-1)^F)^2=(-1)^{F+\bar{F}}$ so that modding out by $\tilde{G}$ yields in this case an orbifold of Type IIB by $(-1)^{F+\bar{F}}$  which is  Type 0B, so that we recover a Type 0B orientifold.}. We are thus lead to the following orientifold projections:

\eject
\noindent 
\centerline{$\underline{\hbox{Type IIB Orientifolds}}$}

\noindent
$G_1=\{1,\Omega\}$

\noindent
$G_2=\{1,\Omega(-1)^{F+{\bar F}}\}$

\noindent 
\centerline{$\underline{\hbox{Type 0B Orientifolds}}$}

\noindent
$G_1=\{1,\Omega\}$

\noindent
$G_2=\{1,\Omega(-1)^{F}\}$

\noindent
$G_3=\{1,\Omega(-1)^{f}\}$

\noindent 
\centerline{$\underline{\hbox{Type 0A Orientifolds}}$}

\noindent
$G_1=\{1,\Omega\}$

Of these orientifolds, we study here the Type IIB orientifolds and the Type 0B string modded out by $G_3$, which suffer from potential anomalies and RR tadpoles. The rest of the models -- which have no anomalies nor  RR tadpoles -- have already been studied in \GomisVI\  (see also 
\lref\BergmanYP{
  O.~Bergman and S.~Hirano,
  ``The cap in the hat: Unoriented 2D strings and matrix(-vector) models,''
  JHEP {\bf 0401}, 043 (2004)
  [arXiv:hep-th/0311068].
}
\BergmanYP) in the context of a non-perturbative description in tems of a matrix model\foot{Type 0B modded out by $G_3$ was not considered in \GomisVI\ since the super-Lioville interaction breaks the $(-1)^f$ symmetry. Here we consider the pure dilaton vacuum state, for which $(-1)^f$ is a symmetry.}.
It would be very interesting to  find a matrix model description of the models studied here.

\newsec{Anomalies and Tadpoles}

The cancellation of anomalies
\lref\AlvarezGaumeIG{
  L.~Alvarez-Gaume and E.~Witten,
  ``Gravitational Anomalies,''
  Nucl.\ Phys.\ B {\bf 234}, 269 (1984).
}
\AlvarezGaumeIG\GreenSG\  is an important constraint on a string vaccum, which has to be diagnosed in our two dimensional superstring theory constructions. 
In two dimensions, a right handed Majorana-Weyl fermion transforming in an $n$-dimensional representation $\rho$ of the gauge group $G$ yields the following anomaly polynomial
\eqn\anomferm{
I_{1/2}(\rho)={n\over 96}\hbox{tr}R^2-{1\over 4}\hbox{tr}_\rho F^2,}
while a right moving   chiral boson contributes:
\eqn\anombose{
I_{C}={1\over 48}\hbox{tr}R^2.}
The spectrum of Type IIB in two dimensions is chiral, but as mentioned in \SeibergBX\ anomalies cancel:
\eqn\anomIIB{
I_{IIB}=2\times I_{1/2}(\rho=1)-I_C=0.}

In the following  we compute the closed string spectrum of the unoriented string theories constructed by modding out by the orientifold groups presented in the previous section. We shall see 
that the closed string spectrum is anomalous. Cancellation of RR tadpoles, however, introduces precisely the right chiral open string degrees of freedom to cancel anomalies. 
 In this section we present the formulas that are needed to compute RR tadpoles, which we are going to apply to the specific models in the rest of the note.

The computation of the tadpole due to the orientifold plane  is easiest to perform by evaluating the Klein bottle vacuum diagram. 
The tadpole
can be obtained by factorizing the loop diagram into the closed tree channel, which encodes the one point function responsible for the tadpole.

 In the two dimensional context under consideration, all particles are massless and the computation of the Klein bottle is easiest to perform as a sum over the target  space {\it closed} string spectrum
\eqn\kb{
Z_{KB}(\Omega g)={V\over 2}\int{dt\over 2t} \int {d^2p\over (2\pi)^2}e^{-\pi t p^2}\sum_{S\in {\cal H}_c} (-1)^{\bf{F}_S}(\Omega g)_S,}
where $(-1)^{\bf{F}_S}=\pm 1$ is the space-time fermion number of the state $S$ and $(\Omega g)_S=\pm 1$ is the eigenvalue of the operator $\Omega g$ on the state $S$. 
The tadpole can be extracted by factorizing to the tree channel
\eqn\kbfact{
Z_{KB}(\Omega g)={V\over 8\pi^3}\int_0^\infty ds \sum_{S\in {\cal H}_c}  (-1)^{\bf{F}_S}(\Omega g)_S,}
where $s=\pi/2t$ for the Klein bottle. 

Whenever an orientifold plane carries charge under a RR 2-form field $C_2$, the tadpole may be cancelled by adding D1-branes charged under $C_2$. Whether tadpole cancellation can be accomplished hinges on whether the contribution from the Klein bottle can be cancelled by the contribution from the M\"obius strip  and cylinder diagram. 
These can be computed by summing over the {\it open} string spectrum on the D1-branes. 
The cylinder diagram on $N$ such D-branes is given by
\eqn\cyl{Z_{C}(\Omega g)=N^2{V\over 2}\int{dt\over 2t} \int {d^2p\over (2\pi)^2}e^{-2\pi t p^2}\sum_{S\in {\cal H}_o} (-1)^{\bf{F}_S},}
which yields in the tree channel
\eqn\cylfact{
Z_{C}(\Omega g)=N^2{V\over 32\pi^3}\int_0^\infty ds \sum_{S\in {\cal H}_o}  (-1)^{\bf{F}_S},}
where $s=\pi/t$ for the cylinder. 

In order to compute the last diagram,  the M\"obius strip, we recall how an orientifold group element acts on an open string state $|S\ra$. This state is represented by an $N\times N$ matrix $S$ and the orientifold projection $G=\{1,\Omega g\}$ keeps the state satisfying
\eqn\acton{
S= (\Omega g)_S \gamma_{\Omega g}S^T\gamma_{\Omega g}^{-1},}
where $(\Omega g)_S=\pm 1$ is the eigenvalue of $\Omega g$ on the state $|S\ra$ and $S^T$ denotes the transpose of the matrix $S$. A universal  open string mode  is a gauge field $A$ for which $(\Omega g)_{A}=-1$ always  so that consistency allows for two possible gauge groups:
\eqn\consist{\eqalign{
\gamma_{\Omega g}=\gamma_{\Omega g}^T\rightarrow &\ SO(N)\ \hbox{gauge group}\cr
\gamma_{\Omega g}=-\gamma_{\Omega g}^T\rightarrow &\ Sp(N/2)\ \hbox{gauge group}.}}
The eigenvalue $(\Omega g)_{S}$ for  the state $|S\ra$ determines  the representation $\rho$ of the  gauge group  under which $S$ transforms.

The M\"obius strip is then given by 
\eqn\mob{
Z_{M}(\Omega g)=\hbox{Tr}(\gamma_{\Omega g}^{-1}\gamma_{\Omega g}^T){V\over 2}\int{dt\over 2t} \int {d^2p\over (2\pi)^2}e^{-2\pi t p^2}\sum_{S\in {\cal H}_o}  (-1)^{\bf{F}_S}(\Omega g)_S,}
which gives in the tree channel
\eqn\mobfact{
Z_{M}(\Omega g)=\hbox{Tr}(\gamma_{\Omega g}^{-1}\gamma_{\Omega g}^T){V\over 8\pi^3}\int_0^\infty ds \sum_{S\in {\cal H}_o}  (-1)^{\bf{F}_S}(\Omega g)_S,} 
where $s=\pi/4t$ for the M\"obius strip.

Therefore, tadpole cancellation in the orientifold by $G=\{1,\Omega g\}$  requires that:
\eqn\tadpole{
N^2\sum_{S\in {\cal H}_o}  (-1)^{\bf{F_S}}+4\sum_{S\in {\cal H}_c}  (-1)^{\bf{F}_S}(\Omega g)_S+4\hbox{Tr}(\gamma_\Omega^{-1}\gamma_\Omega^T)\sum_{S\in {\cal H}_o}  (-1)^{\bf{F}_S}(\Omega g)_S=0.}

We  now show for the various new models defined in section $2$ that tadpoles can be cancelled and that 
the precise chiral degrees of freedom needed to  cancel anomalies arise in the open string sector.

\newsec{New Unoriented Noncritical Strings}

Having already gathered the relevant formulas for anomalies and tadpoles in the previous section, we now just  require computing the action of $\Omega g$ on the various closed and open string states.

The basic symmetry we mod out by is worldsheet parity, generated by $\Omega$. 
In order to fully characterize the models, we must specify the action of $\Omega$ on worldsheet fermions, spin fields and the closed string vacuum state. It is  given by:
\eqn\actionon{\eqalign{
\Omega:&\matrix{&\psi&\rightarrow {\bar\psi}\cr
                                   &{\bar\psi}&\rightarrow -\psi}\Longrightarrow \Omega: {\bar \psi}\psi \rightarrow{\bar \psi}\psi\cr
\Omega:&\matrix{&S^\alpha\rightarrow {\bar S^\alpha}\cr
                                  &{\bar S^\alpha}\rightarrow S^\alpha}\Longrightarrow \Omega: {\bar
S^\alpha}\otimes S^\beta\rightarrow-{\bar
S^\beta}\otimes S^\alpha\cr
&\Omega\cdot |0\ra_{NS-NS}= |0\ra_{NS-NS}.}}

We now study the various string theories.

\subsec{\hbox{Modding Type IIB by} $G_1=\{1,\Omega\}$}

Given the action of $\Omega$ in \actionon\ it follows that the only state surviving the projection is a right-moving Majorana-Weyl fermion $\Psi_-^S$, where
\eqn\defino{
\Psi_-^S\equiv\Psi_-+\tilde{\Psi}_-\qquad \Psi_-^A\equiv\Psi_--\tilde{\Psi}_-,}
while $\Psi_-^A$ and $C_+$ are projected out.

The spectrum of the model, as it stands, is anomalous. Extra degrees of freedom
must be added to the system to cancel gravitational anomalies. We now show that the introduction of the required  degrees of freedom  follows by imposing  cancellation of RR tadpoles.

The Klein bottle diagram \kbfact\ is non-vanishing since
\eqn\specta{
\sum_{S\in {\cal H}_c}(-1)^{\bf{F_S}}(\Omega)_S=-1+1-1=-1,}
and the O1-plane carries RR charge and the model is inconsistent. We can cure this by adding 
suitable D1-branes such that the total RR charge vanishes. 

We need to construct the boundary state  corresponding to a D1-brane in two dimensional Type IIB string theory. The corresponding boundary state is given by
\eqn\boundIIB{
|D1\ra={1\over \sqrt{2}}\left( |+\ra_{NSNS}-|-\ra_{NSNS}\right)+{1\over \sqrt{2}}\left( |+\ra_{RR}-|-\ra_{RR}\right),}
which yields the following GSO projection and spectrum  in the open string channel
\eqn\spect{
\matrix{&\underline{D1-brane}\cr
& \cr
&\hskip-20pt(NS+)\rightarrow A \cr
&(R+)\rightarrow \lambda_+,}}
where $A$ is a non-dynamical gauge field and $\lambda_+$ is a right moving Majorana-Weyl spinor transforming in the
representation $\rho$ of the gauge group $G$.

Using the fact that
\eqn\actionopen{
(\Omega)_{\lambda_+}=-1,} 
we find that $\lambda_+$ transforms in the same way as the gauge field, i.e in the $\rho=$ adjoint representation of the gauge group $G$. We can now compute the cylinder \cylfact\ and M\"obius strip \mobfact\ contribution to the tadpole:
\eqn\tadpoleomega{\eqalign{
\sum_{S\in {\cal H}_o}(-1)^{\bf{F_S}}&=-1\cr
\sum_{S\in {\cal H}_o}(-1)^{\bf{F_S}}(\Omega)_S&=-1\cdot -1=1.}}

Tadpole cancellation \tadpole\ requires that 
\eqn\tadpoleomega{
N^2+4-4\hbox{Tr}(\gamma_{\Omega}^{-1}\gamma_{\Omega}^T)=0,}
which implies that $N=2$ and that $\gamma_{\Omega}=\gamma_{\Omega}^T$.
Therefore, the open string gauge group is $G=SO(2)$ and $\lambda_+$ transforms  in the $\rho$=adjoint representation of $SO(2)$, which for $SO(2)$ is the trivial $\rho=1$ representation.

Summarizing, the spectrum of the model is:
\medskip
\medskip
\noindent
$\underline{\hbox{Spectrum of Type IIB}/\{1,\Omega\}}$
\smallskip

\noindent
$\bullet$ Closed: $\Psi_-^S$

\noindent
$\bullet$ Open: $SO(2)$\ \hbox{gauge field}\ $A$; $\lambda_+$\ \hbox{in $\rho=1$ rep. of $SO(2)$}
\medskip
\medskip

The fermion structure coming from the open string sector and which follows from tadpole cancellation makes the model free of gravitational and gauge anomalies.

\noindent
\subsec{\hbox{Modding Type IIB by} $G_1=\{1,\Omega(-1)^{F+\bar{F}}\}$}

Given the action of $\Omega$ in \actionon\ and the usual action of $(-1)^{F,(\bar{F})}$ (which acts  by -1 on the corresponding left(right) moving spin field)
it follows that the only state surviving the projection is the right-moving Majorana-Weyl fermion $\Psi_-^A$, while $\Psi_-^S$ and $C_+$ are projected out. The closed string spectrum of this model is the same as the previous one, and is therefore anomalous.

The Klein bottle diagram \kbfact\ is non-vanishing  and is the same as before
\eqn\spectb{
\sum_{S\in {\cal H}_c}(-1)^{\bf{F_S}}(\Omega (-1)^{F+\bar{F}})_S=-1+1-1=-1,}
and the O1-plane carries RR charge and the model is inconsistent. We can cure this by adding 
suitable D1-branes such that the total RR charge vanishes. 

The cylinder diagram \cylfact\ is also the same as before
\eqn\cylomegaff{
\sum_{S\in {\cal H}_o}(-1)^{\bf{F_S}}=-1,}
so that the open string spectrum is a  gauge field  $A$ and a right moving Majorana-Weyl spinor  $\lambda_+$ transforming in the
representation $\rho$ of the gauge group $G$.

Using the fact that
\eqn\actionopen{
(\Omega(-1)^{F+\bar{F}})_{\lambda_+}=+1,}
we find that $\lambda_+$ transforms differently than  the gauge field; $\lambda_+$ transforms\foot{For $G=SO(N)(Sp(N/2))$ the adjoint is the antisymmetric(symmetric) representation.} in the $\rho=$ symmetric(antisymmetric) representation of $G=SO(N)(Sp(N/2))$.

The M\"obius strip diagram \mobfact\ is different since $(-1)^{F+\bar{F}}$ acts non-trivially on $\lambda_+$
\eqn\mobomegaff{
\sum_{S\in {\cal H}_o}(-1)^{\bf{F_S}}(\Omega(-1)^{F+\bar{F}})_S=-1\cdot 1=-1.}

Tadpole cancellation \tadpole\  in this model requires that 
\eqn\tadpoleomega{
N^2+4+4\hbox{Tr}(\gamma_{\Omega}^{-1}\gamma_{\Omega}^T)=0,}
which implies that $N=2$ and that $\gamma_{\Omega}=-\gamma_{\Omega}^T$.
Therefore, the open string gauge group is $G=Sp(1)\simeq SU(2)$ and $\lambda_+$ transforms in the 
$\rho=\anti$ representation of $Sp(1)$ which for $Sp(1)$ is the trivial $\rho=1$ representation. 

\medskip
\medskip

Summarizing, the spectrum of the model is:
\medskip
\medskip
\noindent
$\underline{\hbox{Spectrum of Type IIB}/\{1,\Omega(-1)^{F+\bar{F}}\}}$
\smallskip

\noindent
$\bullet$ Closed: $\Psi_-^A$

\noindent
$\bullet$ Open: $Sp(1)$\ \hbox{gauge field}\ $A$; $\lambda_+$\ \hbox{in $\rho=1$ rep. of $Sp(1)$}

\medskip
\medskip

The fermion structure coming from the open string sector and which follows from tadpole cancellation makes the model free of gravitational and gauge anomalies.

\medskip
\medskip

\noindent
\subsec{\hbox{Modding Type 0B by} $G_1=\{1,\Omega(-1)^{f}\}$}

Given the GSO projection of Type 0B string theory \specto\  and the action of $\Omega$ in \actionon\ it follows that 
the right moving scalar $C_-$ is the only state state surviving the orientifold projection while $T$ and $C_+$ are projected out. Since $C_-$ is a chiral field in two dimensions, the model is anomalous. 

We now show that also in this model the constraint imposed by tadpole cancellation cancels all anomalies. 

The Klein bottle diagram \kbfact\ is given by
\eqn\spectc{
\sum_{S\in {\cal H}_c}(-1)^{\bf{F_S}}(\Omega)_S=-1-1+1=-1,}
and the O1-plane carries RR charge and the model is inconsistent. We can cure this by adding 
suitable D1-branes such that the total RR charge vanishes. 

It is well known that in Type 0B string theory (before orientifolding)   there are two different D1-branes, whose boundary states are given by:
\eqn\bound{\eqalign{
|D1,+\ra&=|+\ra_{NSNS}+|+\ra_{RR}\cr 
|D1,-\ra&=|-\ra_{NSNS}+|-\ra_{RR}.}}
The action of $\Omega (-1)^f$ exchanges $|D1,+\ra$ with  $|D1,-\ra$. Therefore in order to be able to mode out by our symmetry $\Omega (-1)^f$, we must have the same number $N$ of $|D1,+\ra$ branes as 
$|D1,-\ra$ branes.

Correspondingly, there are three sectors of open strings corresponding to $D1_+-D1_+$ strings, $D1_--D1_-$ strings and $D1_+-D1_-$ strings (the $D1_--D1_+$ strings are the complex conjugate of the $D1_+-D1_-$ strings). The boundary state overlaps produce the following GSO projection and spectrum in the open string channel:
\medskip
\eqn\spectop{\eqalign{
\matrix{&\underline{D1_+-D1_+} &\underline{D1_--D1_-}&\underline{D1_+-D1_-}\cr
&  & &\cr
&(NS,+)\rightarrow A_1&(NS,+)\rightarrow A_2&(R,+)\rightarrow  \lambda_+}}}
\medskip
\medskip
$A_1$ and $A_2$ are $U(N)$ gauge fields while $\lambda_+$ is a right moving {\it complex} Weyl fermion transforming under the bifundamental representation of $U(N)\times U(N)$. 

After the orientifold projection we are left with a  $U(N)$ gauge field $A$,  since the two gauge groups above get identified by the action of $\Omega (-1)^f$. Furthermore, using that 
\eqn\accioferm{
(\Omega(-1)^f)_{\lambda_+}=-1,}
we find that $\lambda_+$ transforms in $\rho=\anti$   representation of $U(N)$.
The cylinder \cylfact\ and M\"obius strip \mobfact\ are given by:
\eqn\tadpoleomeganew{\eqalign{
\sum_{S\in {\cal H}_o}(-1)^{\bf{F_S}}&=-1\cr
\sum_{S\in {\cal H}_o}(-1)^{\bf{F_S}}(\Omega(-1)^f)_S&=-1\cdot -1=1.}}

Tadpole cancellation \tadpole\ then requires that
\eqn\tadpoletypeo{
N^2+4-4N=0,}
which implies that $N=2$. Therefore, the open string gauge group is $G=U(2)$ and $\lambda_+$ transforms in $\rho=\anti$ representation of $U(2)$, which  for $U(2)$ is the $\rho=1$ trivial representation.

Summarizing, the spectrum of the model is:
\medskip
\medskip

\noindent
$\underline{\hbox{Spectrum of Type 0B}/\{1,\Omega(-1)^f\}}$

\smallskip
\noindent
$\bullet$ Closed: $C_-$

\noindent
$\bullet$ Open: $U(2)$\ \hbox{gauge field}\ $A$; complex $\lambda_+$\ \hbox{in $\rho=1$ rep. of $U(2)$}
\medskip
\medskip

The fermion structure coming from the open string sector and which follows from tadpole cancellation makes the model free of gravitational and gauge anomalies since the anomaly polynomial vanishes $I_C-2\times I_{1/2}(\rho=1)=0$, as shown in \anomIIB.

\bigbreak\bigskip\bigskip\centerline{{\bf Acknowledgements}}\nobreak
\medskip\medskip

We would like to thank Nati Seiberg and Nemani Suryanarayana for discussions. Research at the Perimeter Institute is supported in part by funds from NSERC of Canada and by MEDT of Ontario.

\listrefs

\end

Type IIA string theory is not worldsheet parity invariant. Nevertheless, if we combine worldsheet parity with the action of (euclidean) time reversal $R$, which acts  by $x\rightarrow -x$, the operator $\Omega R$ is a symmetry of the theory. The discrete symmetry group of Type IIA string theory is also given by the dihedral group of four elements, which is now generated by $a=\Omega R (-1)^F$ and $b=(-1)^{F}$.

Using the interpretation of the vacuum diagrams in terms of the sum over the space-time spectrum, we can compute the Klein bottle diagram, which captures the RR tadpole of the O1-plane
\eqn\kb{
Z_{KB}={V\over 2}\int{dt\over 2t} \int {d^2p\over (2\pi)^2}e^{-\pi t p^2}\cdot (-1)=-{V\over 16\pi^2}\int{dt\over t^2}=-{V\over 8\pi^3}\int_0^\infty ds,}
where $s=\pi/2t$ for the Klein bottle. 
The $(-1)$ is due to the contribution of the space-time fermion $\Psi_-^S$, while the contribution of $\Psi_-^A$ is cancelled by $C_+$. 

Therefore, the O1-plane carries RR charge and the model is inconsistent. We can cure this by adding 
suitable D1-branes such that the total RR charge vanishes. 

The open string spectrum on a  D1-brane in two dimensional Type IIB string theory is just a left moving Majorana-Weyl fermion $\lambda_+$, which is obtained by implementing the usual GSO projection:
\eqn\spect{
\matrix{&\underline{D1-brane}\cr
& \cr
&\hskip-20pt(NS+)\cr
&(R+)\rightarrow \lambda_+.}}
The cylinder diagram due to $N$ such D1-branes in the orientifold model is given by:
\eqn\cyl{
Z_{C}=N^2{V\over 2}\int{dt\over 2t} \int {d^2p\over (2\pi)^2}e^{-2\pi t p^2}\cdot (-1)=-N^2{V\over 32\pi^2}\int{dt\over t^2}=-N^2{V\over 32\pi^3}\int_0^\infty ds,}
where $s=\pi/t$ for the Cylinder. 
The $(-1)$ is due to the contribution of the space-time fermion $\lambda_+$.

This shows that the D1-brane also carries RR charge. Whether there is an exactcancellation between the charge of  the O1-plane and that of a configuration of  D1-branes can be found by computing the M\"obius strip diagram:
\eqn\mob{
Z_{M}=\hbox{Tr}(\gamma_\Omega^{-1}\gamma_\Omega^T){V\over 2}\int{dt\over 2t} \int {d^2p\over (2\pi)^2}e^{-2\pi t p^2}\cdot (-1)\cdot (-1)=\hbox{Tr}(\gamma_\Omega^{-1}\gamma_\Omega^T)
{V\over 32\pi^2}\int{dt\over t^2}=\hbox{Tr}(\gamma_\Omega^{-1}\gamma_\Omega^T){V\over 8\pi^3}\int_0^\infty ds,}
where $s=\pi/4t$ for the M\"obius strip. 
The first $(-1)$ is due to the contribution of the space-time fermion $\lambda_+$ while the second 
one is because $\lambda_+$ is odd under the action of $\Omega$. The  $N\times N$ matrix $\gamma_\Omega$ implements the action of $\Omega$ on the D1-brane Chan-Paton factors.

Therefore the total amplitude is given by:
\eqn\total{
-{V\over 32\pi^3}\left(N^2+4-4\hbox{Tr}(\gamma_\Omega^{-1}\gamma_\Omega^T)\right).}
Consistency conditions allow for two possibilities
\eqn\consist{\eqalign{
\gamma_\Omega=\gamma_\Omega^T\rightarrow SO(N)\ \hbox{gauge group}\cr
\gamma_\Omega=-\gamma_\Omega^T\rightarrow Sp(N/2)\ \hbox{gauge group},}}
so that \total\ can be written as 
\eqn\totalfin{
-{V\over 32\pi^3}\left(N\mp2\right)^2.}

Therefore, tadpoles are cancelled if we impose the $O1^-$ projection and choose $N=2$. By doing this we have cancelled the RR tadpole, but we have also added new open string degrees of freedom, namely a non-dynamical $SO(2)$ gauge field and a  left moving Majorana-Weyl fermion $\lambda_+$ transforming in the adjoint representation of the gauge group, so it is a neutral fermion.

The fermion structure coming from the open string sector and which follows from tadpole cancellation makes the model free of gravitational and gauge anomalies. 

\noindent
$\underline{Spectrum\ of\ Type\ IIB/\{1,\Omega(-1)^{F+\bar{F}}\}}$

\noindent
$\bullet$ Closed: $\Psi_-^A$

\noindent
$\bullet$ Open: $\lambda_+$

\newsec{Type IIA Orientifolds}

\end

Using the interpretation of the  Klein bottle, M\"obius strip and Cylinder diagram in terms of space-time vacuum energy and the spectrum of the model
\

The orientifold projection introduces an O1-plane. If the O1-plane carries RR charge, additional D1-branes must be included to have vanishing of total RR charge. Whether a given O1-plane carries RR-charge can be diagnosed by computing the Klein bottle diagram.

The Klein bottle for this orientifold is given by
\eqn\orienKB{
Z_{KB}=\hbox{Tr}_{NSNS,RR}\left({\Omega\over 2}\cdot {1+(-1)^f\over 2} {1+(-1)^{\bar f} \over 2}\right),
}
where $f(\bar{f})$ denotes left(right) moving worldsheet fermion number. 

Given \actionon\ it follows that
\eqn\compute{
\hbox{Tr}_{NSNS,RR}\left(\Omega\cdot {1+(-1)^{f+\bar{f}}\over 2}\right)=0.}
Upon factorizing on the tree channel, this computation shows that the overlap of the crosscap state $|C\ra$ corresponding to the O1-plane with the  NS-NS states vanishes.

The remaining pieces are given by
\eqn\rrf{\eqalign{
&\hbox{Tr}_{NSNS}\left({\Omega\over 4}\cdot {(-1)^f+(-1)^{\bar{f}}\over 2}\right)=-{1\over (2\pi)^3}\cr
&\hbox{Tr}_{RR}\left({\Omega\over 4}\cdot {(-1)^f+(-1)^{\bar{f}}\over 2}\right)=-{1\over (2\pi)^3}.}}
The nonvanishing of this terms shows that there is a non-vanishing overlap between $|C\ra$ and a RR-field, thus yielding a RR tadpole. It is this tadpole that we  want to cancel by suitably adding D1-branes.

Type IIA string theory is not worldsheet parity invariant. Nevertheless, if we combine worldsheet parity with the action of (euclidean) time reversal $R$, which acts  by $x\rightarrow -x$, the operator $\Omega R$ is a symmetry of the theory. The discrete symmetry group is also given by the dihedral group of four elements, which is now generated by $a=\Omega R (-1)^F$ and $b=(-1)^{F}$.

Just like in the Type IIB theory, there are two choices of Type IIA orientifold projection:

\noindent
$G_1=\{1,\Omega R\}$

\noindent
$G_2=\{1,\Omega R(-1)^{F+{\bar F}}\}$

\noindent
$\bullet$ Type 0 Theories
\eqn\spect{\eqalign{
\matrix{&\underline{Type\ 0B} &\underline{Type\ 0A}\cr
&  &\cr
&\hskip-20pt(NS+,NS+)&(NS+,NS+)\cr
&(NS-,NS-)\rightarrow T&\hskip+20pt(NS-,NS-)\rightarrow T\cr
&(R-,R-)\rightarrow C_-&\hskip-20pt(R-,R+)\cr
&\hskip-8pt(R+,R+)\rightarrow C_+&\hskip-20pt (R+,R-)}}}

Therefore, the spectrum of particles of the Type 0B string  is  two scalars $T$  and $C=C_++C_-$, while the Type 0A string has a single scalar field $T$.

The basic symmetry we mod out by is worldsheet parity, generated by $\Omega$. 
In order to fully characterize the model, we must specify the action of $\Omega$ on worldsheet fermions, spin fields and the vacuum state. It is  given by:
\eqn\actionon{\eqalign{
\Omega:&\matrix{&\psi&\rightarrow {\bar\psi}\cr
                                   &{\bar\psi}&\rightarrow -\psi}\Longrightarrow \Omega: {\bar \psi}\psi \rightarrow{\bar \psi}\psi\cr
\Omega:&\matrix{&S^\alpha\rightarrow {\bar S^\alpha}\cr
                                  &{\bar S^\alpha}\rightarrow S^\alpha}\Longrightarrow \Omega: {\bar
S^\alpha}\otimes S^\beta\rightarrow-{\bar
S^\beta}\otimes S^\alpha\cr
&\Omega\cdot |0\ra_{NS-NS}= |0\ra_{NS-NS}.}}

\listrefs

\end

\cdot\psi={\bar \psi},\
\Omega\cdot{\bar\psi}=-{\psi}\Longrightarrow\ 
\Omega\cdot({\bar \psi}\psi)={\bar \psi}\psi,\cr
&\Omega\cdot S^\alpha={\bar S^\alpha},\
\Omega\cdot{\bar S^\alpha}=S^\alpha\Longrightarrow\ \Omega\cdot({\bar
S^\alpha}\otimes S^\beta)=-{\bar
S^\beta}\otimes S^\alpha.}}

The worldsheet description of $D=2$ superstring theory is
given by a  free ${\cal N}=1$ superfield tensored with ${\cal N}=1$
super-Liouville with ${\hat c_L}=9$.  ${\cal N}=1$
super-Liouville is defined by an ${\cal N}=1$ superfield $\Phi$ with
components $(\phi,\psi)$ and 
a superpotential $W=\mu e^{b{\Phi}}$. The GSO projection giving rise to
Type 0B string theory is given by:
\eqn\GSO{\eqalign{
\hbox{NS-NS}:&\ {1\over 2}(1+(-1)^{F+\tilde F})\cr
\hbox{RR}:&\ {1\over 2}(1+(-1)^{F+\tilde F}).}}
The spectrum of the model consists of a massless tachyon $T$ arising form the
NS-NS sector and an additional massless scalar $C$ arising from the RR sector. 

Due to the non-chiral GSO
projection, the theory is
manifestly invariant under the action of worldsheet parity
${\Omega}$. We define the action of ${\Omega}$ on the fermions and
on the spin fields as follows:
\eqn\actionon{\eqalign{
&\Omega\cdot\psi={\bar \psi},\
\Omega\cdot{\bar\psi}=-{\psi}\Longrightarrow\ 
\Omega\cdot({\bar \psi}\psi)={\bar \psi}\psi,\cr
&\Omega\cdot S^\alpha={\bar S^\alpha},\
\Omega\cdot{\bar S^\alpha}=S^\alpha\Longrightarrow\ \Omega\cdot({\bar
S^\alpha}\otimes S^\beta)=-{\bar
S^\beta}\otimes S^\alpha.}}

The action of $\Omega$ can be combined with the action of
other $\ZZ_2$ symmetries to yield other consistent orientifold
projections. The theory is  invariant under $(-1)^{F^L_s}$,
where $F^L_s$ is left-moving space-time fermion number, so that one
can consider  
modding out the theory by the following two orientifold groups\foot{In
ten dimensions one can also consider the model (see e.g.)
obtained by modding out
by the orientifold group
$G=\{1,\Omega(-1)^F\}$. In $D=2$, $(-1)^F$ is not a symmetry since it
flips the sign of the term ${\bar \psi}\psi e^{b\phi}$ in the
super-Liouville action.}:

\eqn\modelos{\eqalign{
\hskip-190pt 1)\;G_1&=\{1,\Omega\}\cr
\hskip-190pt 2)\;G_2&=\{1,\Omega(-1)^{F^L_s}\}.}}

We now analyze the physics of these two models.

\end